\documentclass[dvips]{article}

\usepackage{icrc2011}

\newcommand{\eh}[1]{\,\mathrm{#1}}

\newcommand{\dg}{^{\circ}}

\newcommand{\pcnt}{\eh{\%}}

\newcommand{\mr}[1]{\mathrm{#1}}

\renewcommand{\epsilon}{\varepsilon}

\newcommand{\tin}[1]{_{\mr{#1}}}

\title{Advanced stereoscopic gamma-ray shower analysis with the MAGIC telescopes}
\newcommand{\etal}{\MakeLowercase{\textit{et al. }}} 
\shorttitle{Lombardi \etal Advanced stereoscopic analysis with MAGIC}

\authors{
         Saverio Lombardi$^{1}$, 
         Karsten Berger$^{2}$, 
         Pierre Colin$^{3}$, 
         Alicia Diago Ortega$^{2}$, 
         Stefan Klepser$^{4}$\\
         for the MAGIC collaboration
         }
\afiliations{
             $^1$Universit\`{a} di Padova and INFN, I-35131 Padova, Italy\\
             $^2$Universidad de La Laguna and Instituto de Astrof\'{i}sica de Canarias, E-38200 La Laguna, Spain\\
             $^3$Max-Planck-Institut f\"{u}r Physik, D-80805 M\"{u}nchen, Germany\\
             $^4$IFAE, Edifici Cn., Campus UAB, E-08193 Bellaterra, Spain\\
             }
\email{saverio.lombardi@pd.infn.it}

\abstract{
The MAGIC experiment was upgraded to a two-telescope system in 2009. Unlike 
other Imaging Air Cherenkov Telescope arrays, MAGIC has operated for five 
years exclusively in monoscopic mode, and the single telescope analysis was 
optimized throughout this time. To improve the analysis, we used techniques 
like the random forest event classification method for different purposes, and 
sophisticated image cleaning algorithms. The monoscopic performance was 
optimized in the energy domain around and below 100 GeV, which is inaccessible 
for the other arrays of Cherenkov telescopes. Still, with these analysis techniques, 
we were competitive also in the TeV regime. In the recent development of 
the stereoscopic analysis chain, the know-how of these single telescope techniques 
was combined with the new possibilities of the three-dimensional reconstruction, 
taking advantage both of the richness of single images and their projections onto 
the sky. We present recent advancements in the image cleaning and direction 
reconstruction algorithms, sky mapping and other procedures currently used in 
the analysis of MAGIC stereo data.
}

\begin{document}
\maketitle

\section{Introduction}
The MAGIC (Major Atmospheric Gamma-ray Imaging Cherenkov) experiment for ground-based 
gamma-ray astronomy is a system of two telescopes operating in stereoscopic 
mode since fall 2009 at the Roque de los Muchachos, Canary Island of La Palma 
(28.8$\dg$~N, 17.8$\dg$~W, 2200 m a.s.l). The first telescope, MAGIC-I, has been 
operating since late 2003, whereas the second one, named MAGIC-II, has been successfully 
commissioned during 2009 \cite{cortina2009}. With the start of the operations of the 
stereoscopic system, the standard analysis package of the MAGIC collaboration has been 
upgraded in order to perform the three-dimensional reconstruction of the recorded atmospheric 
showers. The development of the stereoscopic analysis chain took advantages of the know-how 
achieved during the single telescope phase as well as of new analysis algorithms which 
have led to a significant enhancement of the performance of the system \cite{carmona2011}.\\
MARS (MAGIC Analysis and Reconstruction Software) \cite{mars} is the official analysis package of MAGIC,
and is a collection of ROOT-based \cite{root} programs written in C++ for the analysis of 
data from gamma-ray Cherenkov telescopes. The data analysis chain implemented in MARS 
is divided into several steps, each of which is performed by an independent program which 
takes as input the output of one or more of the previous stages \cite{moralejo2009}. 
%
%
\section{Image cleaning}
\begin{figure}[hbt!]
\vspace{5mm}
\centering
\includegraphics[width=3.in]{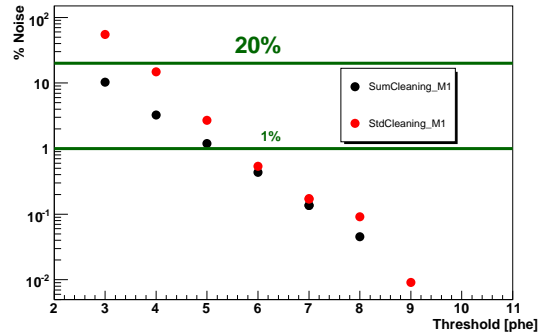}
\caption{
Cleaning level thresholds for the core pixels versus the surviving noise for MAGIC-I data. The noise is estimated 
using pedestal triggers that do not contain Cherenkov light from shower images. The sum cleaning always performs 
better and even at the lowest possible setting of 3~phe an acceptable maximum of 10$\%$ of the noise 
survives in the data.
}
\label{fig1}
\end{figure}

The initial input to MARS are the raw data recorded by the telescopes, consisting of binary files 
containing the full information available per pixel (digitized signal amplitude vs. time) for every 
triggered event, plus ascii files containing regular reports from the different telescope subsystems.
The data are calibrated (separately for each telescope) in order to extract the signal of each pixel 
(after pedestal offset subtraction), its arrival time, and to convert the reconstructed signal amplitudes 
into physically meaningful units (photoelectrons [phe]). For details on the calibration procedure see 
e.g. \cite{moralejo2009}.\\
After the calibration, the data of each telescope are processed in order to remove pixels which most 
likely do not belong to a given shower image (image cleaning procedure) and subsequently to perform 
a parameterization of the resulting cleaned image. The image cleaning is a necessary procedure since 
signals can be induced e.g. by night sky background (NSB) fluctuations or electronic noise. Since these 
noise components are not correlated (contrary to the Cherenkov light of the shower images), they can be 
suppressed by searching for a tight correlation of the signal both in time and in space. The current 
standard image cleaning used in the MAGIC analysis chain \cite{aliu2009} is the so-called
``standard time cleaning''. The algorithm first uses a relatively high signal threshold to search for 
at least two neighboring pixels (so-called core pixels) which belong to the core of the shower (the threshold is 6~phe 
in case of MAGIC-I and 9~phe for MAGIC-II). In this step a time coincidence within 1.5~ns between
core pixels is required. In a second step, adjacent pixels (so-called boundary pixels) 
are allowed to pass into further analysis if a lower signal amplitude of 3~phe (MAGIC-I) and 4.5~phe 
(MAGIC-II) is given. The arrival time of the boundary pixels must be within 4.5~ns of the mean arrival time of core pixels.\\ 
A more complex algorithm (so-called ``sum cleaning''), which is currently being optimized, has been 
introduced in \cite{rissi2009}. In this procedure the signals are clipped in amplitude and all possible 
combinations of 2, 3 or 4 neighboring pixels (2NN, 3NN, 4NN) in the camera are summed up.
If this sum is above a certain threshold and within a sharp ($\sim$1~ns) time interval these pixels are 
considered to belong to the shower image. The clipping ensures that afterpulses (or strong NSB fluctuations)
do not dominate the summed pixels. Finally, the second step of the standard time cleaning is used only for those pixels which survive 
first the sum cleaning, but the signal amplitudes can be relaxed, e.g. for Magic-I only 3~phe for the 
core pixels and 2~phe for the boundary pixels is required (6~phe and 3~phe for Magic-II respectively). 
In this way the analysis threshold can be reduced, which is especially important for moon data where 
the NSB can be several times above the dark time average. Figure \ref{fig1} shows a comparison of the 
surviving noise in MAGIC-I data after applying the two cleaning methods at a given threshold for the first cleaning 
level (core pixels). As shown, the sum cleaning performs better than the standard time cleaning at each threshold.
\\
After the cleaning procedure, each image is finally parametrized by a a small set of parameters which 
describe in a compressed way its orientation, shape and timing properties. Among these quantities 
are the Hillas parameters \cite{hillas1985}, which are basically the moments up to second order of 
the light distribution on the camera, and some time-related image parameters, as the evolution of 
the arrival time of the light along the major axis of the shower image.\\
%
%
\section{Shower parameter reconstruction and Background rejection}
Once the events recorded by each telescope are cleaned and parametrized, the two separate telescope 
data streams are merged in single output files and a first estimation of the basic stereo image parameters is performed. 
In this step the global parameters describing an air shower (e.g. direction, altitude of the shower maximum, 
impact parameters relative to each telescope) are estimated using only the main axis and centroid position 
of both telescope images. This geometrical reconstruction is independent of any Monte Carlo (MC) simulation.\\
The suppression of the unwanted background showers produced by charged cosmic rays is achieved
in MARS by means of a RF multivariate classification method \cite{albert2008}. 
In the standard procedure, the algorithm takes a set of both global shower parameters 
(such as the reconstructed maximum height of the showers) and image parameters of each of the telescopes 
(such as the basic Hillas and timing parameters), as input, and produces one single parameter as an output, 
called \emph{hadronness}, which ranges from 0 to 1. A low value of \emph{hadronness} indicates that
the event is a good gamma candidate. For the learning phase of the gamma/hadron separation procedure, 
the RF method uses as training samples sets of files from MC gamma-rays and 
real MAGIC data. The real data come from observations of a sky region devoid of any gamma-ray source (hence containing
almost exclusively background events).\\
A variant of the DISP method \cite{lessard2001} is used to determine an estimated event direction for each of the telescopes 
separately, via a RF fed with image parameters (including time-related ones). For the DISP, the method uses only a set 
of files from MC gamma-rays as training samples. The final reconstructed direction is computed by combining the direction 
estimates from the two telescopes with the purely geometrical reconstruction. The head-tail ambiguity of the DISP method 
is resolved by taking the two closest solutions (one per telescope). The average of these two provides the final estimated direction.
Once the final direction is estimated the other shower parameters are re-calculated 
taking this improved reconstruction into account. 
Then the energy of the events is estimated for each telescope with a look-up table (LUT) based on the
shower parameters. The LUTs are built from MC gamma-ray simulations by default, but can also be estimated using 
the RF method. The final reconstructed energy is the weighted average of individual telescope 
energies, weighted according to the reconstruction uncertainty associated to the LUTs.\\
Finally, the RF output matrices computed in the previous analysis steps are applied to the real data as well as 
to the MC test sample, and output files with the \emph{hadronness}, DISP, and reconstructed energy parameters
are produced. 
These files are the input for the final stages of the analysis where the flux
calculations, the energy spectrum unfolding procedure, and the light-curve determination for a given source are performed. 
Details on these analysis steps can be found elsewhere \cite{moralejo2009}.
%
%
\section{Skymapping}
\begin{figure*}[!t]
\centerline{\includegraphics[width=3.in]{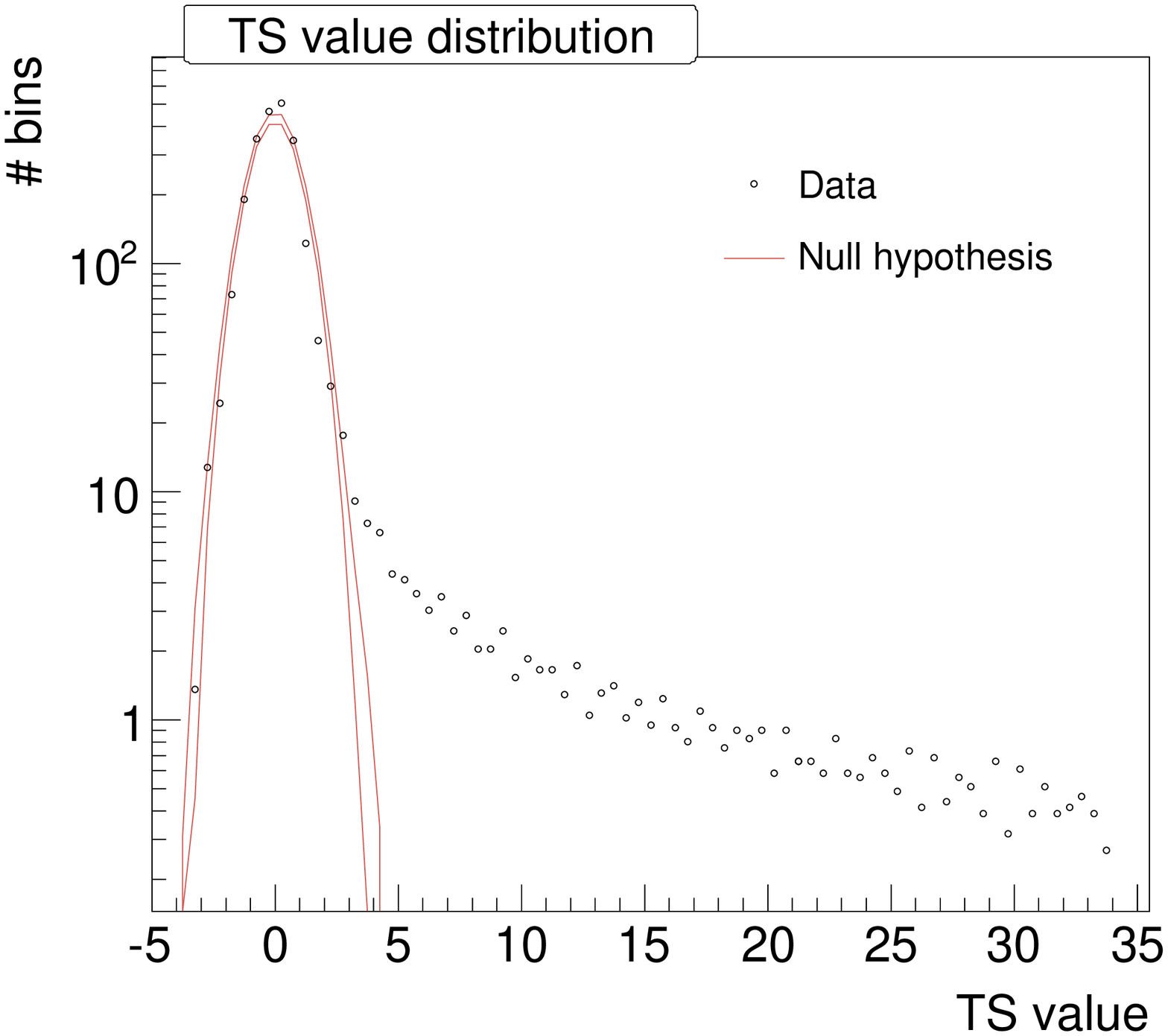}\label{fig2}
            \hfil
            \includegraphics[width=3.in]{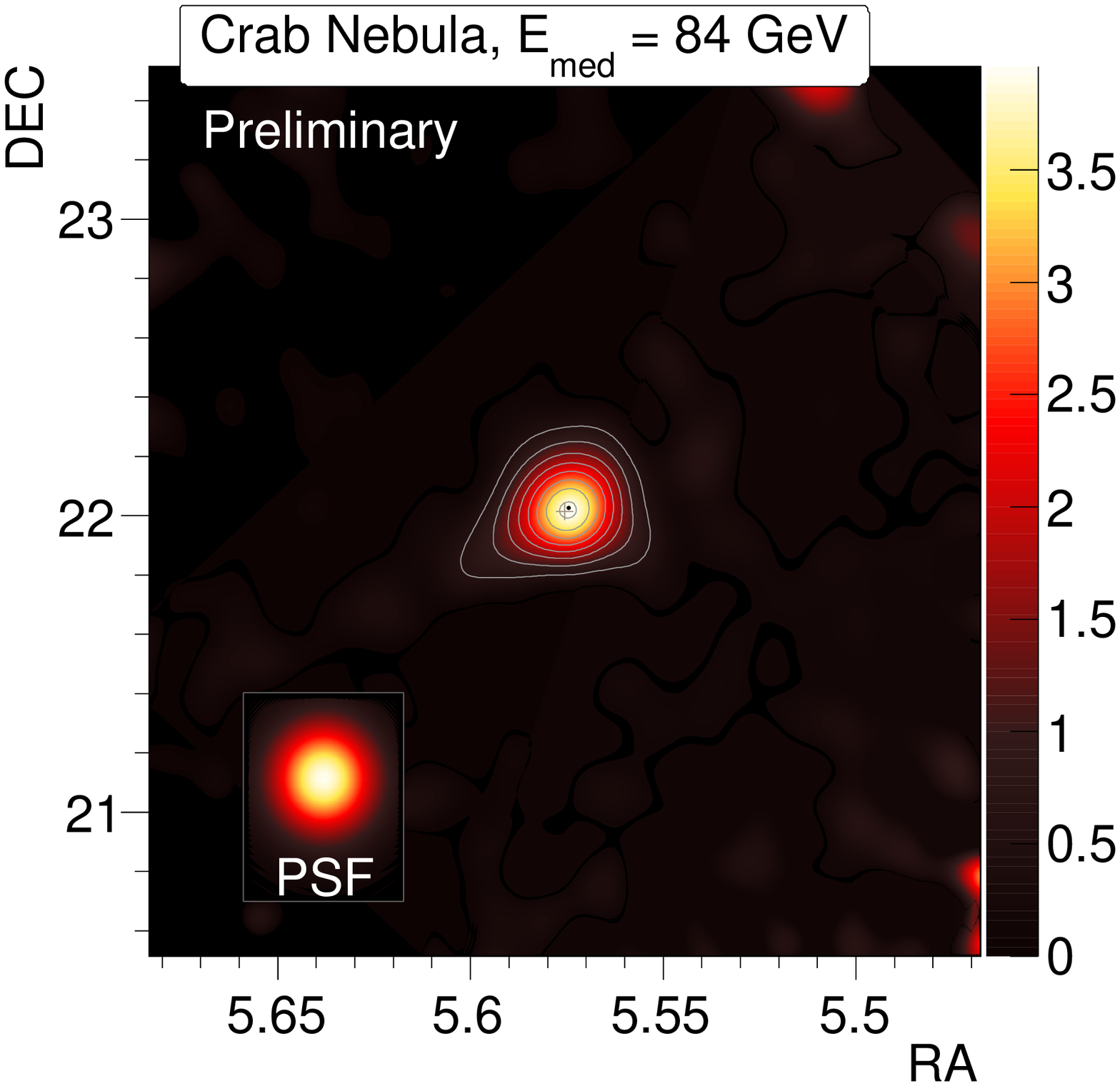} \label{fig3}}
\caption{Left: Distribution of TS values. The null hypothesis distribution closely 
resembles a Gaussian function, so the TS is very similar to a Gaussian significance.\\
Right: MAGIC skymap of the Crab Nebula for events with median true energies
of $84\eh{GeV}$ (see text for description). The relative flux is calculated as described in the
text, and is roughly proportional to the actual gamma flux. Overlaid
are TS value contours in steps of 4, starting at 5. The small black dot is our
fitted Crab Nebula position.}
\label{double_fig}
\end{figure*}

An analysis tool which has been improved for the stereo analysis is the sky mapping procedure. 
The original procedure developed for monoscopic MAGIC analysis was designed to cope both with 
the rapidly inclining off-axis acceptance shape of the MAGIC-I camera, and local exposure inhomogeneities.
Such local acceptance imperfections could be caused by day-to-day electronics or calibration artifacts 
implied by the overall aim to reach the lowest-possible threshold, and are impossible to be modeled in 
detail with simulations. In stereoscopic data, most of these artifacts average out between the two telescopes and are in general 
much less problematic, since the analysis relies more on stereoscopic parameters rather than detailed
shower image features. Instead, however, there are geometrical inhomogeneities, caused by the fact that 
the overlap of the two fields of view of the telescopes is not circularly symmetric, and on top of that 
rotates with the azimuth angle. The basic layout of the skymap analysis was thus kept similarly sophisticated 
to what it was before, but adjusted to stereoscopic analysis in several ways.\\
The present procedure to derive a skymap in MAGIC has three steps:
\begin{enumerate}
\item Construction of a two-dimensional background exposure model (BEM) in relative focal plane coordinates
\item Projection of this model into celestial coordinates, following closely the observed trajectories
\item Comparison of this background model to the actual event distribution with a test statistic, and evaluating 
a relative flux estimator
\end{enumerate}
In wobble mode \cite{fomin1994}, and assuming the source to be in between the two wobble positions, the BEM can simply
be calculated from the photon-like hadron events in the sky areas opposite to the source position in each wobble 
data set. It is formulated in a focal plane coordinate system that is rotated proportionally to the azimuth angle. 
Like this, the orientation of the oval shape of the air shower acceptance is invariant against the azimuth angle. 
Furthermore, the exposure is calculated in bins of the azimuth angle, leading to a three-dimensional BEM.\\
To correctly build a background event expectation map, the BEM is projected to the sky by sampling $N_{\mr{resample}}=200$ 
random events for every photon-like event in the data. For each of these events, the actual pointing position of 
the telescope is used for both the geometrical transformations to sky coordinates, and the choice of BEM azimuth bin. 
A two-dimensional linear interpolation is applied to the BEM of each azimuth bin before the projection to avoid
artifacts from projected bin edges.\\
Before the comparison of the background expectation map with the measured events, a Gaussian kernel density 
folding (smoothing) is applied. The width of this kernel is an analysis parameter to be adjusted depending on 
the point spread function achieved by the individual analysis ($\sigma_{\mr{PSF}}$), the size of the data set, 
and the strength of the source. In a blind search for sources in the field of view, the kernel is chosen to
be $1\,\sigma_{\mr{PSF}}$, a compromise between high resolution and low noise (i.e. trial factor).\\
With the event density expectation map derived in this procedure, we apply the test statistic (TS) defined in \cite{lima} 
with an $\alpha$ that is extremely low ($\approx 1/N_{\mr{resample}}$). In most cases this leads to a Gaussian null hypothesis
distribution, which shows us that the the expectation map, after interpolation, resampling and smoothing, is indeed a very accurate
estimation of the background level. Also, we see that without the above azimuth-related treatments we get strong artifacts 
and significant large-scale biases, so we conclude that these treatments are vital to the skymapping procedure.\\
Still, depending on the amount of events available, and the total exposure at the edge of the sky window, poissonian
components or other effects sometimes lead to slight deviations from a purely Gaussian shape. Therefore we always calculate 
the null hypothesis TS value distribution for every skymap individually by invoking 10 toy simulations of the same sky window with
identical statistical precision, and extract the actual null hypothesis TS distributions from them (see Figure~\ref{double_fig}~(left)).\\
For MAGIC, the physical quantity to display on a skymap is neither the excess events density nor the TS value, because our acceptance and sensitivity are
not flat across the sky window. Instead, we calculate the \textit{relative flux}, defined as the excess events relative
to the background density after smearing  $N\tin{ex}/N\tin{bkg (<0.1\dg)}$. Since the background density (of photon-like hadrons) 
is roughly proportional to the effective area for actual photons, the relative flux is in good approximation proportional to the
absolute flux.\\
As an example skymap at very low energies, Figure~\ref{double_fig}~(right) shows a skymap of the Crab Nebula at estimated
energies below $120\eh{GeV}$. The true energy distribution extracted from MC ranges from $50\eh{GeV}$ to $130\eh{GeV}$ 
($10$ to $90\pcnt$ quantiles, median energy $84\eh{GeV}$). The PSF of the analysis is a roughly 2D gaussian with a sigma
of $0.11\dg$, the smearing kernel used is $0.08\dg$, resulting in a total $\sigma$ of about $0.14\dg$. The shown signal is based on $\approx2000$ excess events.
%
%
%
\section{Summary}
We have presented an overview of the methods implemented in the MARS package for the analysis of data 
from the MAGIC Cherenkov telescopes, with particular emphasis to the latest developed analysis tools 
such as new algorithms for the image cleaning, for the background rejection and direction reconstruction,
and for the sky mapping procedure. The upgrade of MARS after the start of the operations of the stereo system 
has proved to give stable and robust results \cite{carmona2011}: the advanced stereo analysis allows the MAGIC 
telescopes to achieve a sensitivity of (0.76~$\pm$~0.03)$\%$ of the Crab Nebula flux in 50~hr of 
observations in the medium energy range (a factor two better than the one achieved with MAGIC-I alone). 
The angular and energy resolution at those energies are respectively better than 0.07$^{\circ}$ and 16$\%$. 
The gain in sensitivity at lower energies ($<$150~GeV) is even larger (a factor three with respect to the 
monoscopic mode), making MAGIC currently the leading ground-based instrument world-wide in this energy range.
%
%


\clearpage

\end{document}